\documentclass[a4paper]{article}

\usepackage{amsmath}
\usepackage{graphicx}   
\usepackage[american]{babel}
\usepackage[utf8]{inputenc}
\usepackage[T1]{fontenc}
\usepackage[babel=true]{csquotes}
\usepackage[backend=biber,style=ieee, sortcites]{biblatex}
\usepackage{afterpage}
\usepackage{color}		
\usepackage{epsfig}
\usepackage{authblk}
\usepackage{pdfpages}
\usepackage{graphicx,epstopdf}
\usepackage[font={footnotesize}]{caption}

\graphicspath{ {Images/}}

\def\hyp{{\hbox{-}}}
\begin{document}

\title{Evolutionary advantage of a broken symmetry in autocatalytic polymers tentatively explains fundamental properties of DNA }

\author{Hemachander Subramanian$^{1\ast}$, Robert A. Gatenby$^{1,2}$ \\
\normalsize{$^{1}$ Integrated Mathematical Oncology Department,} \\
\normalsize{$^{2}$Cancer Biology and Evolution Program, }\\
\normalsize{H. Lee Moffitt Cancer Center and Research Institute, Tampa, Florida.} \\
\normalsize{$^\ast$To whom correspondence should be addressed; E-mail:  hemachander.subramanian@moffitt.org.} }
\date{\today}
\maketitle

\begin{abstract}
The macromolecules that encode and translate information in living systems, DNA and RNA, exhibit distinctive structural asymmetries, including homochirality or mirror image asymmetry and $3' \hyp 5'$ directionality, that are invariant across all life forms. The evolutionary advantages of these broken symmetries remain unknown. Here we utilize a very simple model of hypothetical self-replicating polymers to show that asymmetric autocatalytic polymers are more successful in self-replication compared to their symmetric counterparts in the Darwinian competition for space and common substrates. This broken-symmetry property, called asymmetric cooperativity, arises with the maximization of a replication potential, where the catalytic influence of inter-strand bonds on their left and right neighbors is unequal. Asymmetric cooperativity also leads to tentative, qualitative and simple evolution-based explanations for a number of other properties of DNA that include four nucleotide alphabet, three nucleotide codons, circular genomes, helicity, anti-parallel double-strand orientation, heteromolecular base-pairing, asymmetric base compositions, and palindromic instability, apart from the structural asymmetries mentioned above. Our model results and tentative explanations are consistent with multiple lines of experimental evidence, which include evidence for the presence of asymmetric cooperativity in DNA. 
\end{abstract}


\section*{Introduction}
Living systems, uniquely in nature, acquire, store and use information autonomously. The molecular carriers of information, DNA and RNA, exhibit a number of distinctive physico-chemical properties that are optimal for information storage and transfer\cite{primitivegeneticpolymers,rnaevolved,rnaevolved2}. This suggests that significant prebiotic evolutionary optimization\cite{primitiveearth1} preceded and resulted in RNA and DNA, and that the nucleotide properties are not simply random.

Here, we concern ourselves with hypothetical self-replicating polymers that evolutionarily preceded DNA and RNA. Specifically, we imagine multiple species of autocatalytic polymers, constructed out of chemically-distinct monomers, competing for common precursors, energetic sources, catalytic surfaces and niches. Our central premise is that the simplest of the evolutionary strategies, higher rates of replication\cite{lifevsprelife}, determined the outcome of this evolutionary competition. We identify some fundamental, common-sense functional requirements that autocatalytic polymers must satisfy in order to replicate faster than other competing species and hence be evolutionarily successful. The goal of this paper is to provide tentative and qualitative explanations for the structural and functional properties of DNA \textit{a posteriori} as adaptations due to the evolutionary pressure to maximize replicational potential.

Evidently, the evolutionary search for the perfect self-replicating molecular species in a given environment is constrained by the diversity of molecules available to be used as monomers in that environment, in the primordial oceans. But, this constraint is intractable, in the absence of well-established knowledge of the chemistry of primordial oceans. We circumvent this biochemical constraint by ignoring its existence, and thus \textit{theoretically assume that evolution was allowed to experiment with an infinite variety of molecular species in its search for the perfectly-adapted monomer}. This assumption translates into freedom for variables and parameters describing the monomers to take on any value, in our mathematical model below. The above premise statement has its roots in the supervenience of evolution over chemistry. Its validation stems from its ability to tentatively explain multiple fundamental properties of DNA, as we will see below.

\section*{The model}
In our simple phenomenological model of a primordial self-replicating system (Methods), we consider an autocatalytic polymer that is capable of replicating without the help of enzymes. A single strand of the polymer catalyzes the formation of another strand on top of itself, by functioning as the template. Free-floating monomers attach to the bound monomers on the template strand at lower temperatures, and facilitate covalent bonding between monomers\cite{andersonTempCycling} and hence polymerization, leading to the formation of the replica strand. The replica strand dissociates from the template strand at higher temperatures, creating two single strands, as happens in Polymerase Chain Reaction. 

A self-replicating molecular species must satisfy certain requirements in order to be evolutionarily successful and to function as an information-carrier. In the following, we list those physically meaningful requirements to be satisfied by the molecular species, and in doing so, arrive at two conflicting requirements. Breaking of a symmetry, upon maximization of replication potential, leads to resolution of the conflict and to simultaneous satisfaction of the two requirements. These requirements are not new, and have been included and explored individually in other models and systems elsewhere\cite{andersonTempCycling,model_like_mine,similartomine,Toymagneticmodel}. 



Self-replication involves both bond formation between free-floating monomers and monomers on the template strand, and bond-breaking between monomers on the two strands, requiring these inter-strand bonds to be relatively weak compared to other bonds in the polymer. On the other hand, information storage requires stronger intra-strand bonds that withstand strong environmental variations, as pointed out by Schr{\"o}dinger\cite{whatislife}. Hence, the self-replicating polymer needs to be composed of two complementary components, mutable inter-strand ``hydrogen bonds'' and relatively immutable intra-strand ``covalent bonds''\cite{andersonTempCycling,model_like_mine,similartomine,Toymagneticmodel}.

The intrinsic covalent bonding rates among free-floating monomers should be lower than the covalent bonding rates between the monomers hydrogen-bonded to the template strand, so that monomers become available for self-replication and not for \textit{de novo} strand formation. This requirement makes self-replication viable and information transfer across generations possible. Evolution could have solved this by identifying monomers whose kinetic barrier for covalent bonding between themselves is lowered when they are attached to the template strand\cite{andersonTempCycling,weak_hydrogenbonds,model_like_mine}. We term this barrier reduction ``\textit{covalent bond catalysis}''.

If a hydrogen bond catalyzed the formation (and hence dissociation as well) of another hydrogen bond in its neighborhood\cite{tidalcycling2}, the strand would be replicationally more successful, since covalent bond formation requires two contiguous monomers hydrogen-bonded to the template. Also, higher rate of monomer attachment to the template would allow for more monomers to be drawn in for polymerization, away from other competing processes such as dimerization through hydrogen bonding. Thus, reduction of kinetic barrier for hydrogen bond formation would be advantageous for the self-replicating system. The foregoing justifies the need for ``\textit{hydrogen bond cooperativity}'', catalysis of hydrogen bond formation/dissociation by their neighboring hydrogen bonds\cite{anticooperative_bonding,polandsheragamodel,PBDmodel,basepairingphysics}. Cooperativity in DNA, the increasing ease of hydrogen bonding between unbonded monomers (zippering) when two single DNA strands are already hydrogen-bonded at one of the ends, is a very well-established phenomenon, and has been well-studied both experimentally and theoretically\cite{basepairingphysics}. The experimental signature of cooperativity in DNA melting is the sharpness of the melting transition, where the DNA goes from a double strand to two single strands within a narrow range of temperature\cite{sharpmeltingdna}. Cooperativity in DNA has also been abundantly documented in DNA zipping and unzipping experiments\cite{sequencedependence1999,cooperativePNAS,cooperativitypnas2,cooperativityPRL}.

Obviously, the probability for the covalent bond formation between two contiguous monomers on the replica strand will increase with the lifetime of the hydrogen bonds of the monomers with the template strand. Thus, higher the kinetic barrier for hydrogen bond dissociation, higher the probability for the successful formation of the covalent bond and hence the replica strand. Thus, we notice that, \textit{while covalent bond catalysis requires higher kinetic barrier for hydrogen bond dissociation, hydrogen bond cooperativity requires lower kinetic barrier for hydrogen bond formation}. Since self-replication requires the replicating polymer to be at or near the melting point of the hydrogen bonds, the kinetic barriers for formation and dissociation are nearly equal, and we arrive at the competing requirement of both higher and lower kinetic barrier height, or equivalently, to fine-tuning of the hydrogen bond lifetime. We could solve this conundrum by introducing an environment with oscillating ambient temperature, where, the hydrogen bond lifetime is longer at lower temperatures and thus enables covalent bond formation, whereas, higher temperatures facilitate strand separation. Nevertheless, strands that \textit{intrinsically} satisfy these two competing requirements would still be evolutionarily more successful, by being able to colonize regions with temperature oscillations of much smaller amplitude.

The solution that simultaneously and intrinsically satisfies these two competing requirements is to break the symmetry\cite{moreisdifferent} of the catalytic influence of a hydrogen-bonded monomer-pair on its two neighboring hydrogen bonds on either side. The hydrogen-bonded monomer-pair can reduce the kinetic barrier for hydrogen bond formation/dissociation to its right, while increasing the barrier for hydrogen bond formation/dissociation to its left, (or vice versa) which we call ``asymmetric hydrogen bond cooperativity''. This solution is similar in spirit to Kittel's single-ended zipper model for DNA\cite{Kittelsmodel}. Asymmetric cooperativity has also been proposed earlier to explain other biophysical processes\cite{asymmetriccooperativity}. Such an arrangement would prolong the lifetimes of the already-formed hydrogen bonds to the pair's left, and thus would increase the probability for covalent bonding among those bonded monomers. It will also enable rapid extension of the replica strand to the right, drawing monomers away from competing processes, by allowing monomers to hydrogen bond with the template easily through the reduction of the kinetic barrier. Thus, the broken symmetry of unequal and non-reciprocal catalytic influence leads to simultaneous satisfaction of the above-mentioned two competing requirements. Surprisingly, the replicational advantage of strands with asymmetric cooperativity over symmetric strands turns out to be crucial for understanding various physico-chemical properties of the extant heteropolymer, DNA.

Our model (methods) simply translates the foregoing in mathematical language. We imagine the construction of a replica strand of an autocatalytic polymer on top of the template strand as a Markov Chain. A Markov chain description of a random process involves identification of the state space, and writing down the transition rates or probabilities between the identified states. Given the transition rate matrix, we can calculate variables that are relevant for our analysis, such as the average first passage time of a given state and average residence time in a given state (methods). We measure the potential of a molecular species to form a replica strand as the product of two factors: the relative rate of monomer utilization for replica strand formation against other competing processes, and the probability for covalent bond formation between any two monomers on the replica strand. The first factor increases with reduction in hydrogen-bonding kinetic barrier, whereas the second factor decreases with the reduction in the barrier height. Asymmetric cooperativity simultaneously satisfies both the requirements, as we show in the next section.

It is crucial to understand that our goal for building this model is limited to demonstrating, with minimal assumptions and in a physically transparent manner, the superiority of primordial self-replicating polymers with asymmetric cooperativity over polymers with symmetric cooperativity in attracting enough monomers to construct the replica strand. In particular, we do not intend for this model to make quantitative predictions about the kinetics of DNA (un)zipping or helix-coil transition, for which, highly sophisticated models already exist\cite{dnastructurefunction}. In keeping with this limited goal, we have included in the model only ingredients that have a direct bearing on our goal. We exclude all other ingredients that provide negligible or no discriminatory capability, even though they might make the model more realistic and accurately reflective of the self-replication process. The ingredients that we reasoned to have the same effect on the self-replication of both symmetric and asymmetric polymers, and is thus non-discriminatory, such as polymer bending, secondary structure formation, multiple monomer types, inclusion of N-mers and so on were thus excluded. In particular, we ignore the differences in the rates of unzipping, between the symmetric and asymmetric-cooperative versions of the double strand, during the high temperature phase of the temperature cycle, by assuming that the time period of the temperature cycles are much larger than the zipping and unzipping times of the double strand, in order to keep the model as simple as possible. This assumption minimizes the contribution of unzipping rates to replication potential, allowing us to solely concentrate on the competition between symmetric and asymmetric polymers for monomers. 

\section*{Methods}
Our aim here is to encapsulate in mathematical language the sequence of hydrogen and covalent bonding and unbinding events that result in self-replication of the autocatalytic polymer. We assume a circular or linear template polymer, constructed by stringing together $N$ monomers through covalent bonding. Free-floating monomers can either hydrogen-bond with each other, forming dimers, at a rate $r_f$, or can bind with the template to initiate the construction of the replica strand, at a rate $r_{t0}$. We denote the presence or absence of a hydrogen bond between a monomer in the replica strand and the $i$-th monomer on the template strand with a $1$ or $0$ in the $i$-th place in a binary string of $N$ digits. Thus, for $N=5$, the binary string $00000$ would imply that the template strand has no monomers hydrogen-bonded to it, and $00100$ implies one monomer hydrogen-bonded to the third monomer on the template strand. Cooperativity of hydrogen bonding is implemented by stipulating different rates for subsequent monomer binding events, depending upon the presence or absence of neighboring hydrogen bonds. The rates $\mathcal{R}$, of monomers hydrogen bonding with template strand in different hydrogen-bonding neighborhoods can then be expressed as
\begin{equation}
\begin{aligned}
& \mathcal{R}\left ( 00000 \rightarrow 00100 \right ) = r_{t0}, \\
& \mathcal{R}\left ( 00100 \rightarrow 00110 \right ) = r_{tr} = \alpha_R r_{t0}, \\
& \mathcal{R}\left ( 00100 \rightarrow 01100 \right ) = r_{tl} = \alpha_L r_{t0} \qquad \text{and}\\
& \mathcal{R}\left ( 01010 \rightarrow 01110 \right ) = r_{tc} = \alpha_R \alpha_L r_{t0}. \\
\label{ratesbonding}
\end{aligned}
\end{equation}

The unbinding rates are

\begin{equation}
\begin{aligned}
& \mathcal{R}\left ( 00000 \leftarrow 00100 \right ) = s_{t0}, \\
& \mathcal{R}\left ( 00100 \leftarrow 00110 \right ) = s_{tr} = \alpha_R s_{t0}, \\
& \mathcal{R}\left ( 00100 \leftarrow 01100 \right ) = s_{tl} = \alpha_L s_{t0} \qquad \text{and}\\
& \mathcal{R}\left ( 01010 \leftarrow 01110 \right ) = s_{tc} = \alpha_R \alpha_L s_{t0}. \\
\label{ratesunbinding}
\end{aligned}
\end{equation}
In Eqs. \ref{ratesbonding} and \ref{ratesunbinding}, $\alpha_R$ and $\alpha_L$ are the factors that modify the rates of hydrogen bonds forming to the right and left of a single hydrogen bond. Symmetric cooperativity results when $\alpha_L=\alpha_R$, and when these two factors are unequal, asymmetric cooperativity results. If we assume that only nearest neighbor hydrogen bonds affect the rate of bonding of another monomer to the template strand, the above rates of bond formation and dissociation are sufficient to determine the rates of transition between all $2^5=32$ states that describe the $N=5$ double-strand formation process.

The rate constants for the transition between all possible states describing the double-strand formation are determined by just four parameters $r_{t0}$, $s_{t0}$, $\alpha_R$, and $\alpha_L$. We analyze part of the self-replication process as a continuous-time Markov Chain process. We can evaluate the time it would take for the template strand to go from one without any monomers attached to it, to one with all of its monomers hydrogen-bonded to a monomer, i.e., from the state $00000$ to $11111$. This is calculated using the well-established ``first passage time'' or ``hitting time'' analysis\cite{stochasticprocesses}: Let $\mathcal{R}_{ij}$ be the transition rate constant from state $i$ to $j$, and $t_i$ the average time taken for the Markov chain to reach the final state $k=11111$ when it begins at state $i$. Then, first passage time analysis involves solving the following set of linear equations for non-negative $t_i$'s:
\begin{equation}
\sum_j \mathcal{R}_{ij} (t_j - t_i) = -1, \qquad i \ne k.
\label{hittingtimeeq}
\end{equation} 
The average time taken to traverse from $00000$ to $11111$, $t_1$ in Eq. \ref{hittingtimeeq}, is henceforth called the ``growth time'' $t_g$. The ``rate advantage'', a measure of the propensity for monomers to hydrogen-bond with a template as opposed to hydrogen-bonding among themselves, is 
\begin{equation}
P_g = \frac{1/t_g}{1/t_g + r_f},
\label{hittingtime}
\end{equation}
where, $r_f$ is the rate of dimerization of monomers.

Let the rate of covalent bond formation between two contiguous monomers attached to the template strand be $r_c$. The probability for the covalent bond to form within a certain time $t$ is then
\begin{equation}
P_c = 1-exp \left ( - r_c t\right ).
\label{probcov}
\end{equation}
The average lifetime of the configuration of a pair of contiguous monomers hydrogen-bonded to the template strand determines the probability of a covalent bond forming between the two monomers, through the above Eq. \ref{probcov}. The most conservative estimate of such a lifetime is the lifetime of the state $11111$, because the last covalent bond has the least time to form, since all other pairs of monomers have been in existence before the last pair. The average lifetime of the state $11111$ is just $1/\mathcal{R}_{11111}$, the inverse of the diagonal entry corresponding to the state $11111$ in the transition rate matrix. The expression for $P_c$ then becomes
\begin{equation}
P_c = 1-exp \left ( - \frac{r_c}{\mathcal{R}_{11111}}\right ).
\label{probcov2}
\end{equation}
  
	While low barrier height for bonding decreases the ``first passage time'' $t_g$ and thus increases the rate advantage $P_g$, it would decrease the covalent bonding time $P_c$. Both fast ``growth time'' (measured by $P_g$) and successful covalent bonding (measured by $P_c$) are important for the success of a self-replicating polymer in creating a full replica strand, which can be measured using the dimensionless metric $P = P_g P_c$, called ``replicational potential'' in this paper. But the conflicting requirements for both these metrics to maximize their respective values, with $P_g$ maximization requiring $\alpha_L, \alpha_R < 1$, and $P_c$ maximization requiring $\alpha_L, \alpha_R > 1$ , sets up a conflict. The conflict is resolved when the left-right symmetry is broken upon maximization of the replicational potential, with $\alpha_L <1$ and $\alpha_R > 1$ or $\alpha_L >1$ and $\alpha_R < 1$. 
	
	\subsubsection*{Parametrization}
	We use the experimentally determined values of the rate constants of hydrogen-bonding, unbinding and covalent bonding among nucleotides on a template strand for evaluating the replicational potential $P$, below. We use the values of $7.5 \times 10^7 M^{-1} sec^{-1}$ and $2.3 \times 10^5 M^{-1} sec^{-1}$, measured at pH $7$ and $269 K$, for the second order rate constants of hydrogen bonding and unbinding, i.e., $r_{t0}$ and $s_{t0}$ respectively, between a nucleotide monomer and a monomer on a template\cite{hydrogenbonding2}. The rates at which monomers attach to and detach from the polymer template is given by the product of the above second-order rate constants and the monomer concentration in the primordial environment. We set that concentration to be $1 nM$, low enough to better bring out the competition between symmetric and asymmetric polymers for monomer substrates. The rate of extension of the replica strand, through covalent bonding between two activated nucleotides hydrogen-bonded to the template, $r_c$, has been measured to be of the order of $10^{-2} min^{-1}$ at pH $8.9$ and $283 K$\cite{incorporationkinetics}. Since our goal is to just demonstrate the replicational superiority of asymmetric polymers, and not quantitative predictions, we ignore the difference in the types of nucleotides used and the values of the environmental variables (pH and temperature) between the above two experiments. In any case, the model is relatively insensitive to the precise values used for the parameters. For simplicity, the rate of dimerization through hydrogen-bonding of two monomers, $r_f$, is taken to be the same as the rate of monomer attaching to the polymer template, $r_{t0}$. The dimensionless catalytic/inhibitory factors $\alpha_L$ and $\alpha_R$ are allowed to independently vary between $0.2$ and $2$\cite{incorporationkinetics}, allowing us to continually interpolate between the symmetric Ising-type interactions and the asymmetric Zipper-type interactions.

\section*{Results}

\begin{figure}
\begin{center}
\includegraphics[width=0.9\textwidth]{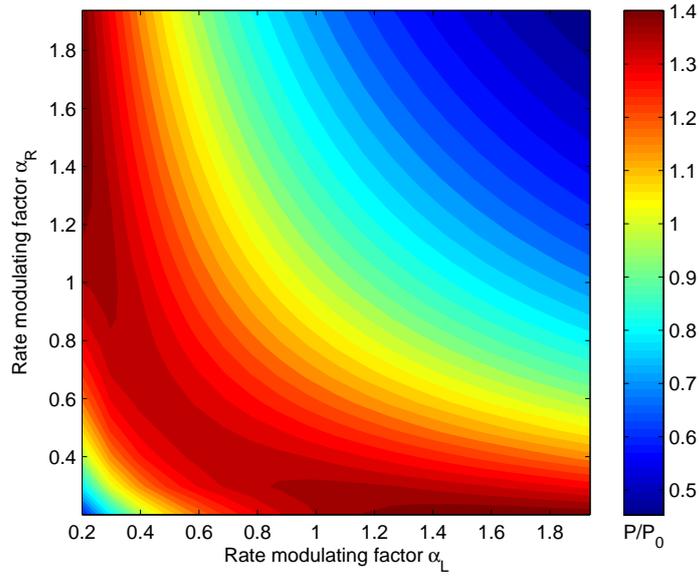}
\caption{Replicational potential $P$ of circular self-replicating polymer strands of length $N=5$, as a function of factors $\alpha_L$ and $\alpha_R$ modulating the rates of hydrogen bonding/unbinding to the left and right of a single pre-existing hydrogen bond. Maximum replicational potential is achieved when the bonding rate of the left bond is reduced as much as possible, and the bonding rate of the right bond is increased above the uncatalyzed rate (or vice versa), resulting in the breaking of left-right symmetry. The two equivalent maxima in the figure, where $\alpha_L \ne \alpha_R$, correspond to two equally possible modes of asymmetric cooperativity. This symmetry-breaking is the consequence of a compromise between two competing requirements for successful self-replication: rapid hydrogen-bonding and unbinding of monomers with the template to speed up replication, and successful formation of covalent bonds between two contiguous hydrogen-bonded monomers, which depend on long hydrogen-bond lifetimes. The replicational potential $P$ above is measured in units of $P_0= P(\alpha_L=1, \alpha_R=1) = 0.04$, the replicational potential without hydrogen-bond cooperativity. The factors that go into calculation of $P$, $P_g$ and $P_c$, are shown in figs. \ref{monomerutil} and \ref{covprob}.}
\label{symmetrybreaking}
\end{center}
\end{figure}

In this section, we intend to prove that the competing requirements mentioned above for evolutionary success in self-replication lead to breaking of the symmetry of catalytic influence of a hydrogen bond on its neighbors on either side. Figure \ref{symmetrybreaking} shows the replication potential $P$ as a function of two variables $\alpha_L$ and $\alpha_R$, the catalytic/inhibitory factors modulating the bonding rates of hydrogen bonds to the left and right of a single pre-existing hydrogen bond. The plot shows two maxima, both equally off the diagonal where the bonding rates are equal, proving our assertion above. This is a \textit{genuine} symmetry-breaking, since \textit{two equivalent degenerate maxima} are present on either side of the symmetric cases (along the diagonal from lower left to top right in fig. \ref{symmetrybreaking}), and both solutions are equally probable. The role played by energy minimization in symmetry-breaking in non-living systems is played here by evolution, i.e., replicational potential maximization. The catalytic arrangement that maximizes the replicational potential is as follows. The right neighbor's barrier is raised as much as possible above its uncatalyzed height and the left neighbor's barrier lowered appropriately (or vice versa), such that the combined effect of lowering and raising the barrier of the central hydrogen bond in return is to slightly lower it below its uncatalyzed height. This symmetry-broken solution is quite insensitive to the values of parameters used, as long as the conflict of requirement for both high and low kinetic barriers remain in effect.  

\begin{figure}
\begin{center}
\includegraphics[width=0.9\textwidth]{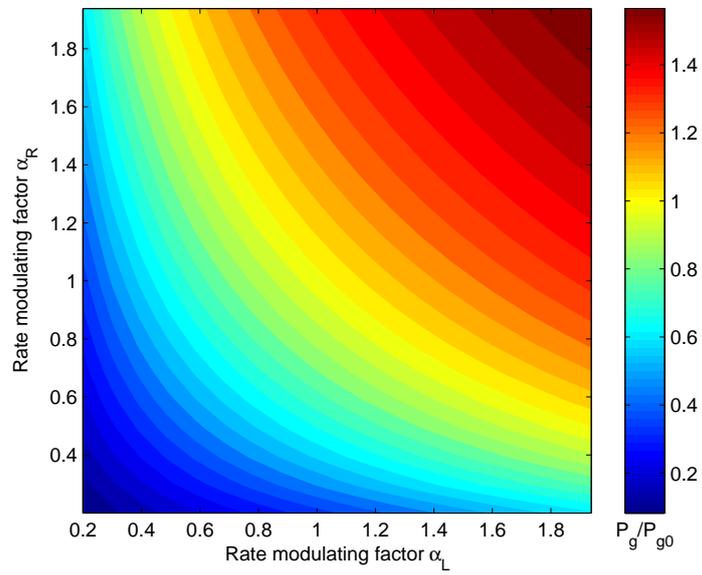}
\caption{Rate advantage $P_g$, normalized with respect to that of a strand with no hydrogen bond cooperativity, as a function of rate-modulating factors $\alpha_L$ and $\alpha_R$. More monomers can be drawn in for template-directed polymerization if the rate of hydrogen-bonding of monomers with the template is high. The maximum rate advantage occurs at the highest possible values of the bonding rates $r_{tl}$ and $r_{tr}$, and where $r_{tl}=r_{tr}$ or $\alpha_L=\alpha_R$.}
\label{monomerutil}
\end{center}
\end{figure}

\begin{figure}
\begin{center}
\includegraphics[width=0.9\textwidth]{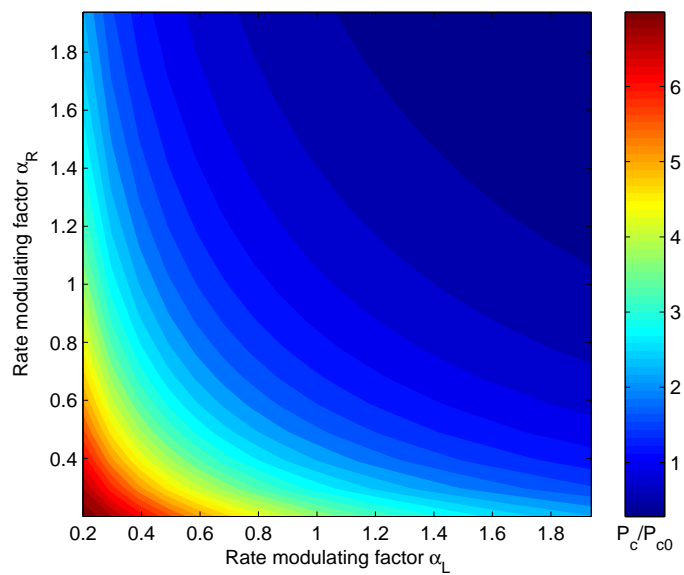}
\caption{Covalent bonding probability $P_c$, normalized with respect to that of a strand with no hydrogen bond cooperativity, as a function of rate-modulating factors $\alpha_L$ and $\alpha_R$. Long lifetime of a pair of contiguous hydrogen bonds increase the covalent bonding probability between the two monomers. High $P_c$ requires low unbinding rates and hence high kinetic barriers near hydrogen-bond melting point. Maximum of covalent bonding probability occurs at the lowest possible values of the bonding rates $r_{tl}$ and $r_{tr}$, and where $r_{tl}=r_{tr}$ or $\alpha_L=\alpha_R$.}
\label{covprob}
\end{center}
\end{figure}

As we mentioned earlier, the replicational potential $P$ in Fig.  \ref{symmetrybreaking} is the product of two factors:  (1) The rate of monomer utilization for polymerization relative to the combined rate of all processes requiring the monomers $P_g$. This relative rate depends upon the rate of monomers hydrogen bonding with monomers on the template strand. The lower the effective barrier height for hydrogen bonding, higher will be the rate of monomer utilization. This is illustrated in Fig.  \ref{monomerutil}, which shows that higher bonding rates of the left and right neighbors lead to higher utilization, and maximum utilization occurs when both left and right rates are \textit{equal}. (2) The probability for covalent bonding $P_c$. This depends on the average lifetime of two contiguous hydrogen bonds, and, higher the barrier height for hydrogen bond dissociation, higher the probability for covalent bonding. This is illustrated in Fig.  \ref{covprob}, where the probability is seen to be high for lower rates of hydrogen bonding, and when both the left and right bonding rates are \textit{equal}. 

Figs. \ref{monomerutil} and \ref{covprob} show that the two factors $P_g$ and $P_c$, whose product is the replicational potential $P$, cannot be simutaneously maximized, since they conflictingly require high and low hydrogen bonding rates, for their respective maximizations. Fig.  \ref{symmetrybreaking} illustrates the compromise between the two requirements for successful self-replication. The compromise that satisfies both the requirements is arrived at by raising the kinetic barrier for hydrogen-bonding to the right and lowering the kinetic barrier for bonding to the left (or vice versa). Thus the replicational potential maxima happen where the bonding rates of the hydrogen bonds to the left and right of a single pre-existing hydrogen bond are \textit{unequal}. This broken symmetry solution provides explanations for multiple fundamental properties of DNA, as we describe below.

\begin{figure}
\begin{center}
\includegraphics[width=0.9\textwidth]{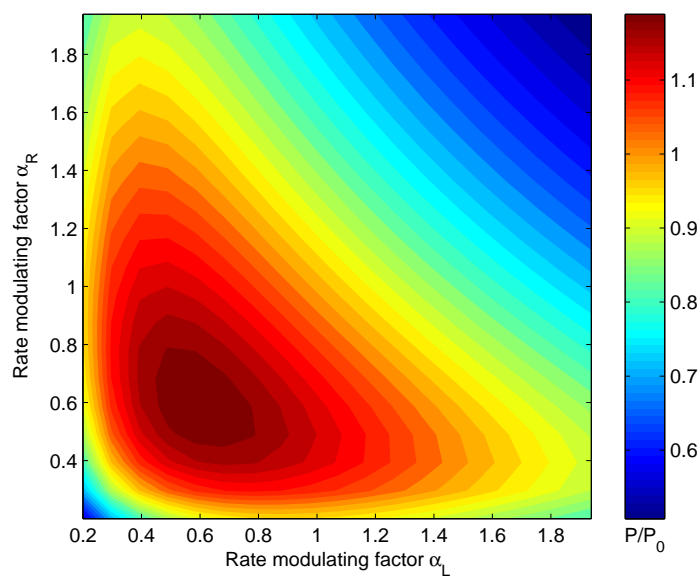}
\caption{Replicational potential $P$ of linear self-replicating polymer strands as a function of factors $\alpha_L$ and $\alpha_R$ modulating the rates of hydrogen bonding to the left and right of a single pre-existing hydrogen bond. Maximum replicational potential occurs when the bonding rates of left and right bonds are equal, $r_{tl}=r_{tr}$. This maximum value is lower than the maximum values in the circular strand case, demonstrating the replicational superiority of circular strands over linear strands. The replicational potential $P$ above is measured in units of $P_0= P(\alpha_L=1, \alpha_R=1)= 0.04$. Broken symmetry compromise for the two competing requirements of self-replication is unviable in linear strands due to the high kinetic barrier in one direction (needed for covalent bond formation) inhibiting hydrogen bonding.}
\label{linstrand}
\end{center}
\end{figure}

Interestingly, only circular strands adopt the above broken-symmetry solution, to satisfy the two competing requirements. The maximal replicational potential of linear strands is lower than the cicular strands' maxima, and occurs where the bonding rates of left and right hydrogen bonds are equal, i.e., when the strands are symmetrically cooperative, as shown in Fig.  \ref{linstrand}. The reason behind the difference between circular and linear strand behavior is as follows. In the circular strand case, the first hydrogen bond connecting a free-floating monomer and a monomer on the template strand can form at any monomer position on the strand, whereas, in the linear strand case, the first hydrogen bond must form only at one of the ends, in order for the strand to self-replicate as effectively as the circular strand. Since asymmetric cooperativity increases the barrier for hydrogen bond formation to the right, formation of the first hydrogen bond at any location other than the right-most template monomer will result in severe inhibition of bond formation to the right of that first bond (replacing ``right'' with ``left'' results in an equally valid statement). This reduces the effectiveness of self-replication, and thus disincentivizes the adoption of asymmetric cooperativity as a solution for satisfying the two competing requirements, in linear strands. 

\clearpage

\section*{Experimental support for asymmetric cooperativity}
The \textit{central prediction} of our model above is the presence of asymmetric cooperativity in evolutionarily successful self-replicating polymers, which includes DNA. Asymmetric cooperativity, unequal catalysis of hydrogen bonds on the left and right, can manifest itself  by \textit{rendering kinetics of zipping of two DNA single-strands and unzipping of DNA double-strands from the left and the right ends, unequal}. This is an eminently experimentally observable phenomenon.

We first point to the evidence for the presence of \textit{directional asymmetry} (deferring differentiating between its thermodynamic and kinetic origins to the following paragraphs), the inequivalence of left and right side in DNA, because of it being a well-established fact in Biology. Let us denote the base-pairing of nucleotides (the four different types of monomers present in DNA) on the top and the bottom strands of the double-stranded DNA as $5' \hyp X \hyp 3'/3' \hyp Y \hyp 5'$, with $5' \hyp X \hyp 3'$ in the top strand hydrogen-bonded to $3' \hyp Y \hyp 5'$ in the bottom strand. The growth of a replica strand on a single strand template DNA happens only in one direction, and the numbers $5'$ and $3'$ are used to identify that direction. Directional asymmetry in DNA can be easily demonstrated by using the well-established nearest-neighbor \textit{thermodynamic} parameters of DNA\cite{ThermodynamicAsymmetry}, wherein, the free energy, enthalpy and entropy of different combinations of nearest-neighbor pairs were experimentally measured and cross-verified. It can be seen from the tables in \cite{ThermodynamicAsymmetry} that, adding $5' \hyp G \hyp 3'/3' \hyp C \hyp 5'$ base-pair to the \textit{left} of  $5' \hyp A \hyp 3'/3' \hyp T \hyp 5'$, resulting in $5' \hyp GA \hyp 3'/3' \hyp CT \hyp 5'$, and adding the same $5' \hyp G \hyp 3'/3' \hyp C \hyp 5'$ to the \textit{right} of $5' \hyp A \hyp 3'/3' \hyp T \hyp 5'$, resulting in $5' \hyp AG \hyp 3'/3' \hyp TC \hyp 5'$, are different operations, result in distinct chemical structures, and obviously have different nearest-neighbor \textit{thermodynamic} parameters. Thus our asymmetric cooperativity prediction merely extends such asymmetric thermodynamic influence to \textit{kinetics}. A note on terminology: The inter-strand bonding between the nucleotides $A$ and $T$ is composed of two hydrogen bonds, and between $G$ and $C$, three hydrogen bonds. Since we have no need to distinguish between either of the two hydrogen bonds between $A$ and $T$ or between the three bonds between $G$ and $C$, we collectively refer the bonds between $A$ and $T$, and between $G$ and $C$ in singular, as ``a hydrogen bond''. Thus, ``interactions between neighboring hydrogen bonds'' would imply interaction between hydrogen bonds of two neighboring base-pairs, and not between the hydrogen bonds of a single base-pair.

A crucial piece of evidence for the existence of directional asymmetry in the \textit{kinetics} of DNA, i.e., asymmetric cooperativity, comes from studying the incorporation kinetics of activated nucleotides that nonenzymatically extend a primer attached to a template strand, one nucleotide at a time, in the presence of a downstream binding strand\cite{incorporationkinetics}. First, the rate of incorporation of a nucleotide is shown to be dependent on the type of nucleotide present on the $3'$ and the $5'$ neighboring ends of the incorporated nucleotide (table 1 of \cite{incorporationkinetics}). Second, the rate of incorporation depends on the orientation of the neighboring base-pairs, i.e., $5' \hyp G \hyp 3'/3' \hyp C \hyp 5'$ versus $5' \hyp C \hyp 3'/3' \hyp G \hyp 5'$. For example, $5' \hyp C \hyp 3'/3' \hyp G \hyp 5'$ supports higher rate of nuceotide incorporation to its left compared to $5' \hyp G \hyp 3'/3' \hyp C \hyp 5'$, whereas $5' \hyp G \hyp 3'/3' \hyp C \hyp 5'$ supports higher incorporation rate to its right compared to $5' \hyp C \hyp 3'/3' \hyp G \hyp 5'$ (Fig. S6 in the supporting information of \cite{incorporationkinetics}). Third, the direction of asymmetric enhancement ($5' \hyp C \hyp 3'/3' \hyp G \hyp 5'$ catalyzing the \textit{left} neighbor) of the incorporation rate agrees with the direction of catalysis that we arrived at from the well-established relationship between the direction of replication and $GC$ skew (please see ``Heteromolecular base-pairing and asymmetric nucleotide composition'').

More experimental evidence for asymmetric cooperativity come in part from unzipping experiments. In one experiment\cite{lambdadnaevidence}, a single-molecule phage $\lambda$ DNA is unzipped using force applied on a microscopic glass slide attached to it. The measured forces of unzipping from one end is shown to be different from unzipping from the opposite end, and this is explained by the group as due to the presence of stick-slip motion\cite{lambdadnaevidencefull}. The fact that different forces signatures are needed to unzip the DNA molecule from either end implies that the work done to unzip the DNA from either end is also different. Under the near-equilibrium experimental conditions of unzipping as mentioned in the article\cite{lambdadnaevidence}, this difference in the unzipping forces cannot be due to thermodynamics. Thus the difference can only be due to the difference in kinetics of unzipping from either end, strongly supporting the presence of asymmetric cooperativity in DNA.

In another set of experiments\cite{directionalasymmetryjacs,jacssequencedependence}, the average unzipping times for a single molecule double-stranded DNA were found to be different depending upon the strand orientation during entry of the strand into the nanopore. This result is explained in those papers using the analogy of  a ``christmas tree'' moving through a hole, with the asymmetry of kinetics arising from the asymmetry of the tree structure. These experiments demonstrate the directionally asymmetric response of base-pair lifetime to nanopore probe. It is thus not unreasonable to assume that the bonding state of the left and right neighboring hydrogen bonds could similarly influence the lifetime of the middle hydrogen bond asymmetrically. In another experiment, even though the thermodynamic stabilities of the two sequences $5' \hyp(AT)_6(GC)_6\hyp 3'$ and $5' \hyp(GC)_6(AT)_6\hyp 3'$ are nearly the same, their unzipping kinetics have been shown to differ by orders of magnitude\cite{sequenceorderunzip}, suggesting that thermodynamics alone cannot explain the sequence functionality, and directionally asymmetric kinetic influences must be included. Unzipping kinetics of a DNA hairpin has been shown\cite{orientationdiscrimination} to strongly depend upon orientation of the terminal base-pairs. In another experiment\cite{duplexzippering}, adding the same four-nucleotide sequence to the $5'$ end of a longer sequence and to the $3'$ end of the same sequence resulted in significantly different zippering kinetics, with the effects on kinetics due to secondary structure formation explicitly ruled out.

In experiments on Prokaryotic adaptive immune systems (CRISPR), the bacterial RNA transcript that attacks the homologous sequence in the DNA genome of an invading bacteriophage has been shown to bind to the latter through an unidirectional zippering mechanism\cite{zippingunidirectional2,unidirectionalzipping1}. A seven-nucleotide subsequence at the $5'$ end of the CRISPR RNA transcript first binds to its complementary subsequence at the $3'$ end of the phage DNA target sequence. The rest of the RNA sequence then zippers along the rest of the DNA to complete the binding process. This unidirectional RNA-DNA zippering provides more evidence for our asymmetric cooperativity model. The zipping directionality has also been demonstrated in a DNA \textit{triple}-helix system\cite{triplehelixdirectional}, where the third strand is observed to nucleate and zip along the double-strand template only in the $5' \hyp 3'$ direction. Other indicators for the presence of asymmetry in DNA include differences in the kinetics of unzipping, depending upon whether the DNA is stretched from $3' \hyp 3'$ or from $5' \hyp 5'$\cite{stretchingdirectionality,stretchingdirectionality2}. Transport studies have also noted the effects of directional asymmetry on charge transfer in DNA\cite{chargetransferasymmetry,chargetransferasymmetry2,electrontransportdirectionality}. Wobble base-pairs have been shown to influence the kinetics of the neighboring base-pairs asymmetrically, destabilizing the neighboring base-pair on only one side\cite{wobbleasymmetry}. Even though the sequences on either side of a DNA base-pair or a lesion is symmetric, the destabilizing effect of the base-pair on the neighboring base-pairs was found to be asymmetric\cite{palindromeevidence,lesionasymmetry}. Kinetics of DNA-protein binding has been shown to depend asymmetrically on the neighborhood, with the sequence flanking the binding region on one side affecting the kinetics more than the other side\cite{tataasymmetry,longrangeinteractionpromoters}. The experiments cited above strongly suggest the presence of asymmetric cooperativity in DNA, which makes perfect sense, given the evolutionary advantage it provides to autocatalytic heteropolymers.

Further support for our hypothesis of asymmetric cooperativity in DNA comes from its ability to explain multiple intriguing observations about DNA sequence and function that have no evolution-based explanation so far, within an unified framework. These observations include the apparent inefficiency of the multi-step lagging strand replication mechanism, the presence of asymmetric nucleotide composition that reduces coding potential, molecular homochirality that prevents life from utilizing enantiomers with opposite chirality in a racemic molecular mixture, and the choice of four nucleotide alphabets instead of two. Local symmetrization, leading to a reduction in asymmetric cooperativity and hence the stability of the double-strand, unifyingly explains all of the following:  a) the propensity for both $AT$ -rich sequences and CpG islands to function as replicational and transcriptional origins, b) the instability of palindromes, inverted repeats and trinucleotide tandem repeats without invoking various secondary structures, c) the skewed distribution of dinucleotide sequences within a stretch of DNA replicated in the same direction, and d) the sequence-dependent local melting of double-strand DNA under negative superhelical stress.

\subsubsection*{An experimental prediction}
Here, we make an experimentally verfiable claim, which cannot be explained by, to our knowledge, the only model explicitly built to explain the differences in the unzipping rates of DNA from either ends\cite{lambdadnaevidencefull}. Within the picture we developed here, the rates of unzipping of the sequence $5' \hyp (C)_n \hyp 3'/3' \hyp (G)_n \hyp 5'$, at constant force, should be different depending on the end where unzipping begins, and the rate of unzipping from the left end should be faster than from the right end, as suggested by both the experiment on incorporation kinetics\cite{incorporationkinetics} and genomic studies on $GC$ skew\cite{SkewReviewRocha,anotherskewreview}. This hypothesized outcome cannot be explained by the model constructed in \cite{lambdadnaevidence,lambdadnaevidencefull}, since that model requires sequence asymmetry to explain unzipping asymmetry, whereas, the above sequence is homogeneous.

\section*{Asymmetric cooperativity tentatively explains fundamental properties of DNA}
We have demonstrated above that asymmetrically cooperative self-replicating polymers stood a great chance of winning the evolutionary competition for resources, due to their high replicational potential. Here, we describe the implications of that conclusion for the extant autocatalytic heteropolymer, DNA, assuming that the precursor of DNA did undergo the directional symmetry-breaking in cooperativity and bequeathed it to the DNA. The experimental evidence listed above strongly suggests that to be the case. Following tentative explanations for diverse characteristics of DNA are based on the supervenience of biochemical evolution over chemistry, as argued in our premise statement, and thus enlarge on the familiar chemistry-based rationalizations.

Below, we explain various properties of DNA, such as strand directionality, circularity of primitive genomes, anti-parallel strand orientation, heteromolecular base-pairing, asymmetric nucleotide distribution across strands, four-nucleotide alphabet and three-nucleotide codons, palindromic instability, helicity, and chirality. We arrive at the explanations through our assumption of the presence of asymmetric cooperativity in DNA, and by assuming that DNA evolved to maximize speed of replication and to acquire the capability to store information. We point to experimental support for our explanations where available. The tentative explanations for the above properties are illustrated in fig.\ref{explanations}.

We provide a very short introduction to the replication mechanism of DNA here, just enough to understand our arguments and explanations below. Double-stranded DNA is composed of two complementary single strands hydrogen-bonded to each other. During replication, these hydrogen bonds are broken and the double strand is converted into two single strands.  These DNA single strands replicate directionally, with the ``replicational machinery'' that constructs the replica strand moving unidirectionally along the track of a single template strand. This directionality of replication is governed by the directionality of the DNA single strand itself, which is a consequence of the two ends of the strands (and hence of the constituent monomers) being inequivalent. These chemically inequivalent ends (of both the strand and the monomer) are differentiated in Biological literature as the $3'$ and $5'$ ends. The two complementary strands of a double-stranded DNA are aligned anti-parallel, with one strand oriented $5'$ to $3'$, and the other oriented $3'$ to $5'$. During replication, the ``unzipping machinery'' creates a Y-shaped fork, where the double strand is unzipped into two single strands, while moving along the double strand unidirectionally. In contrast to the directional movement of the ``replicational machinery'' on the single strands, the direction in which the ``unzipping machinery'' moves is not dictated by the orientation of either of the two strands in the double strand, but is usually argued to be dictated by ``initial condictions'' during its induction. Since the two strands in the double strand are oriented anti-parallel, the unzipping and replicational machineries move in the same direction on only one of the  two strands, called the leading strand. On the other strand, called the lagging strand, the two machineries move in opposite directions, which results in the the lagging strand replication to happen in fragments. Elaborate cellular mechanisms exist to accurately replicate the inherently risky lagging strand replication.

\subsubsection*{Strand directionality}

We have mentioned above the restriction imposed by the single strand on the direction the ``replicational machinery'' must move in, during the replica strand synthesis. This directionality constraint is so strong and inviolable that, even in the situation where constructing the replica strand in the opposite direction could be continuous and less error-prone, as in the lagging strand replication, the evolutionarily selected molecular structure of DNA disallows such a course\cite{directionalitymodel}. Thus, there must be a strong evolutionary reason for retaining the directional nature of the strand and hence of the replicational machinery, and to forego the obvious advantages of allowing for a change in the direction of the replicational machinery (also see ``Anti-parallel strands'' below). This property is usually explained using chemistry-based arguments, which is unacceptable within our premise of the supervenience of biochemical evolution over chemistry. The obvious question is why would evolution choose a directional monomer and strand structure, when a non-directional strand could have allowed for changes in directionality of replication, when the demand arises. Asymmetric cooperativity readily provides an explanation for the evolutionary advantage of strand directionality, and helps us reconcile the apparently elaborate and involved mechanism of lagging strand replication of DNA, with the DNA's evolutionary superiority.

Asymmetric cooperativity results when two contiguous hydrogen bonds influence each other asymmetrically or non-reciprocally, with the bond on the right, say, reducing the kinetic barrier of its left neighbor, whereas the latter increasing the kinetic barrier of the former. Such an asymmetric influence needs an asymmetric molecule to instantiate it. A molecule with symmetric left and right sides, with the covalent-bonding ends to the left and right structurally symmetric, cannot instantiate asymmetric cooperativity. This leads us to conclude that the monomers must themselves be \textit{reflection-asymmetric} along the covalent-bonding direction. Thus asymmetric cooperativity would lead to monomer and strand directionality. The evolutionary advantage of unidirectional replication due to asymmetric cooperativity thus helps us understand the inefficient lagging strand replication mechanism as a work-around to replicate without foregoing the advantage of asymmetric cooperativity.

\afterpage{%
\thispagestyle{empty}
\begin{figure}[htbp]
\centering
\includegraphics[width=1.2\textwidth]{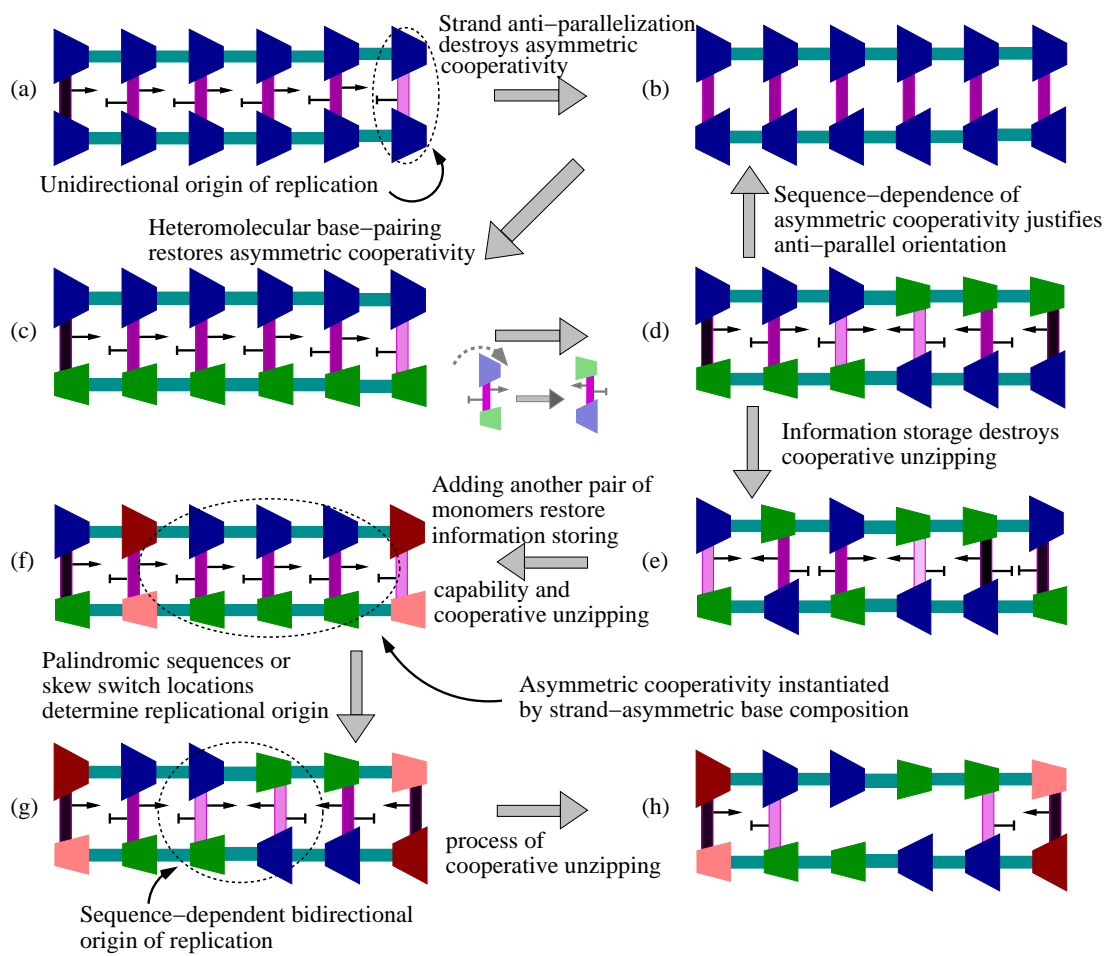}
\caption{Explanations for various properties of DNA, within the asymmetric cooperativity picture. (a) Asymmetric cooperativity requires the monomers constituting a heteropolymer to have asymmetric structure along the covalent bonding direction, represented here as blue trapezoids with unequal opposite side lengths. Asymmetric kinetic influence of two neighboring hydrogen bonds on each other is represented using catalytic and inhibitory arrows drawn on the hydrogen bonds. Darker shading of the bonds denote higher kinetic barriers. The single strand thus acquires directionality, explaining the $3'\hyp5'$ directionality of DNA strands. We have assumed that the linear template and replica strands are oriented in parallel, giving the double strand an overall (un)zipping directionality. (b) Anti-parallel orientation of the two strands destroys overall directionality of the double strand. The two-fold rotationally symmetry of the base-pairs precludes differentiating between left and right side of the base-pairs, illustrating the lack of (un)zipping directionality. (c) Heteromolecular base-pairing reintroduces left-right asymmetry, and restores (un)zipping directionality in the double-strand. One of the two strands dominate and sets the asymmetric cooperativity mode. (d) Rotation of the base-pair by $180^\circ$ leads to switching of catalytic and inhibitory arrows, and thus of (un)zipping directionality. This explains the sequence dependence of (un)zipping directionality, and provides the evolutionary rationale for anti-parallel double-strand orientation. With sequence dependence of asymmetric cooperativity, simultaneous replication of disjoint segments is made possible. (e) To be able to store information, heteropolymers of arbitrary sequences should have similar replicational potential.  Heteropolymers with aribtrary sequences would not exhibit cooperative (un)zipping, due to the dependence of (un)zipping cooperativity on base-pair orientation. Thus, heteropolymers made of just two alphabets cannot store information, since a few sequences will become replicationally superior to a vast majority of other sequences. (f) With the choice of quadruplet alphabet, monomer sequences can simultaneously store information and determine (un)zipping directionality, thus unlinking their mutual influence. Asymmetric base composition sets the asymmetric cooperativity mode and thus the unzipping direction. (g)  Palindromic sequences or $GC$ skew switching locations are dyadic-symmetric monomer arrangements, and indicate a switch in (un)zipping directionality. Symmetric catalytic influence due to dyadic symmetry leads to weakening of hydrogen bonds and to double-strand instability, and thus these locations can serve as origins of replication and transcription. (h) Illustration of bidirectional sequential unzipping originating from the region of dyadic symmetry. This also explains palindromic instability.}
\label{explanations}
\end{figure}
\clearpage
}

\subsubsection*{Circularity of primitive genomes}

 The genomes of most prokaryotes (cells without nuclear membrane), plasmids or extra-chromosomal DNA, and of many viruses, are circular. The evolutionary advantage of a circular genome is posited to be the simplicity of its replication mechanism, as opposed to that of a linear genome, whose free ends require a more sophisticated mechanism for faithful replication. It has also been argued that circular genomes enable storage of negative super-helicity, which favors local unzipping or bubble formation\cite{kamenetskiihelicity}. In the ``results'' section, we have demonstrated another advantage of having a circular genome: Asymmetric cooperativity is beneficial for circular genome, but not for linear genome. This is because of the difficulty in replicating the linear template strand in the direction (say, right) in which hydrogen-bonding is inhibited, which leaves the template region to the right of the origin of replication un-replicated. The origin of replication for a linear strand must be at its right-most end for the strand to successfully replicate itself, a constraint not present in the case of circular genomes. The replicational machinery of the circular genome moving to the left of origin of replication will end up circling the entire genome. Thus, in the primordial oceans, those self-replicating polymers that incorporated asymmetric cooperativity to be replicationally superior were probably circular. This could then provide another explanation for the circularity of genomes of evolutionarily primitive species mentioned above.


\subsubsection*{Anti-parallel strands and multiple origins of replication}

 As we have seen in the explanation for ``strand directionality'', the lagging strand construction is apparently inefficient, because it happens in fragments that are stitched together later. Also, part of the lagging strand is left in a single-stranded state, which is vulnerable to degradation from other cellular components. Apart from the constraint of directionality, the anti-parallel strand orientation is directly responsible for such a cumbersome replicational process. If the DNA strands were parallel, both the strands could be constructed continuously. Naturally, there must be a \textit{strong evolutionary reason} for the DNA to persist with its anti-parallel orientation, as with directionality, to tolerate such an inefficiency. The evolutionary advantage of anti-parallel strand orientation arises from the possibility it offers to \textit{replicate non-overlapping segments of DNA simultaneously}, thereby speeding up the replication process tremendously, as we expain below.

As the Fig.  \ref{symmetrybreaking} illustrates, there are two degenerate replicational maxima representing two possible modes of asymmetric cooperativity: ``Left asymmetric cooperativity'', where the left hydrogen bond's barrier is lowered and the right bond's barrier raised, and ``right asymmetric cooperativity'', where the left hydrogen bond's barrier is raised and the right bond's barrier lowered. Parallel orientation of the two strands of DNA restricts the asymmetric cooperativity to be in either left or right mode, uniformly, throughout the entire genome. Across the entire genome, the hydrogen bonds must catalyze their left neighboring bonds and inhibit their right neighbors (or vice versa), due to the same direction of the monomers in both strands that instantiate asymmetric cooperativity. Thus, the ``unzipping machinery'' can move only in the right direction across the entire genome, during replication. The replication of the entire genome thus has to be done sequentially (see below). With anti-parallel orientation, the two strands negate each other's asymmetric influence (see Fig. \ref{explanations}(b)), and the directionality of the ``unzipping machinery'' is left to other more malleable determinants, which can be made to depend on the local sequence. If one of the strands in the anti-parallel configuration is allowed to \textit{locally} dominate the other strand and determine the mode of asymmetric cooperativity, then the ``unzipping machinery'' could be made to move along the direction dictated by that dominating strand. We will later argue that this is indeed the case.

Now, the directionality of the ``unzipping machinery '' and thus of replication can be altered at will, by locally switching the ``\textit{dominating factor}'' (please see the explanation for ``heteromolecular base-pairing'') from one strand to another, unlike in the case of parallel strands. Segments of DNA can thus be replicated simultaneously, parallelizing the replication process. The places where the strands change their cooperativity mode from left-asymmetric to right-asymmetric is where replication would begin, with a pair of unzipping machineries inducted to move from the ``origin-of-replication'' to the left and right. Multiple such origins of replication would allow for simultaneous unzipping of multiple double-strand segments at once, which results in faster replication time. Because the cooperativity mode is hard-coded into parallel strands, such switching between left and right modes is not possible in parallel-strand organization. This switching between cooperativity modes in anti-parallel strands, achieved through switching of the thus far vaguely defined ``dominating factor'', renders a pair of hydrogen bonds at the switching location to be weaker than their neighbors (the rates of their bonding/unbinding increased by a factor of $\approx 1.5^2$ within our model), due to the absence of inhibitory kinetic influence, as shown in Fig. \ref{explanations}(g)-(h). \textit{These weakened bonds serve as ``origins of replication'' in the anti-parallel strands}. Due to the absence of switching between cooperativity modes, these weakened hydrogen bonds cannot be present in parallel strands, except at the ends, thus preventing the strands from supporting multiple origins of replication.

The presence of simutaneously replicating non-overlapping segments would also enable expansion of genome size and thus the complexity of organisms, by disentangling genome size and replication time. More over, an evolutionary transition from circular to linear strands might not have been possible with parallel strands, due to the instability of the hydrogen bond with smaller kinetic barrier at one of the free ends. Whereas, due to the possibility of cooperativity mode switching, antiparallel strands can strengthen their free ends, by modulating the asymmetry direction just at the ends to make the bonds near the end to have higher kinetic barrier. Evolutionary preference for anti-parallel strand orientation of the double strands may have resulted in the selection of a heteropolymer structure for DNA which is thermodynamically more stable in anti-parallel orientation\cite{parallelstrandlessstable}.

\subsubsection*{Heteromolecular base-pairing and asymmetric nucleotide composition}

 It is not possible to incorporate asymmetric cooperativity in anti-parallel strands with interstrand hydrogen bonds between same type of monomer molecules. To prove this assertion, let us consider a monomer of type $G$ in the top strand hydrogen bonded with another monomer of same type $G$ in the bottom strand. We represent this base-pair as $5' \hyp G \hyp 3'/3' \hyp G \hyp 5'$, with the numbers denoting the anti-parallel orientations of the monomers in the two strands. This base-pair lacks left-right asymmetry, as can be easily demonstrated by rotating it by $180 ^{\circ}$, effectively converting the top strand monomer into the bottom strand monomer and vice versa, which results in the same base-pair arrangement $5' \hyp G \hyp 3'/3' \hyp G \hyp 5'$. Thus, to break the left-right symmetry, we have to abandon the assumption of homomolecular base-pairing, and dictate that base-pairing can happen only between monomers of different types, such as $5' \hyp G \hyp 3'/3' \hyp C \hyp 5'$. This latter heteromolecular base-pairing arrangement is evidently left-right asymmetric, and hence can incorporate asymmetric cooperativity. If $5' \hyp C \hyp 3'/3' \hyp G \hyp 5'$ is left-asymmetrically cooperative, (i.e., catalyze the left hydrogen bond) then $5' \hyp G \hyp 3'/3' \hyp C \hyp 5'$ must be right-asymmetrically cooperative, flipping the direction of catalytic influence, as illustrated in Fig. \ref{explanations}(c)-(d).

In the part explaining the anti-parallel orientation of the two strands of DNA above, we have mentioned about the ``dominating factor'' that dictates which of the two strands gets to choose the direction in which the unzipping machinery must move, or in other words, the mode of asymmetric cooperativity. Here we will clarify what this factor is, and supply evidence for its presence in DNA. In the above paragraph, we explained how the orientation of a heteromolecular base-pair, by which we mean which monomer of the pair is part of which strand, determined the asymmetric cooperativity mode. Let us consider a DNA segment where the unzipping machinery is obsereved to move from left to right. This implies that the strand is locally left-asymmetrically cooperative, which would be instantiated by the base-pairs $5' \hyp C \hyp 3'/3' \hyp G \hyp 5'$, in alignment with the observation that $G$ is enriched in the leading strand\cite{anotherskewreview}. Switching the orientation of this base-pair by placing $G$ in the top strand will change this cooperativity mode and thus would not be preferred. Thus we should observe an excess of $C$ over $G$ in the top strand (lagging strand for left-to-right replication) with corresponding excess of $G$ over $C$ in the bottom strand (leading strand for left-to-right replication), over the entire segment where the unzipping machinery is observed to move from left to right, and vice versa. Thus, the ``dominating factor'' determining the direction of movement of the unzipping machinery is the \textit{relative excess of one nucleotide}, of the two base-pairing nucleotides $G$ and $C$, over another, in a single strand.

This correlation between excess of one nucleotide over another on a strand and the direction of movement of the unzipping machinery is experimentally very well-documented and is observed in the DNA of both prokaryotes and eukaryotes\cite{SkewReviewRocha,anotherskewreview,yeastreplicationorigins,skeworigin,compositionalasymmetryeukaryotes,replicationoriginsmammals}. The excess of one nucleotide over another in a segment of DNA has been termed ``asymmetric base composition'' or ``$GC$ skew'', and the above-mentioned correlation is in fact one of the primary experimental signatures used to identify the multiple origins of replication in various genomes\cite{orilocpaper,origindetection,skeworigin,strandasymmetryPNAS,bacterialreplicationstart}. This strand asymmetry, calculated as $(C-G)/(C+G) \% $ in running windows along genomic sequences, varies from an average of about $4\% $ in Human genome \cite{strandasymmetryPNAS} to more than $12\% $ in some Bacteria\cite{lobryasymmetric}. ``Origins of replication'', mentioned above as locations where the asymmetric cooperativity mode changes between left and right, are correlated with the locations where the $GC$ skew changes sign. These are locations where the dyadic symmetry is approximately restored, with more $G$'s on the top strand to the left of the origin decreasing the kinetic barrier of the bonds at the origin, and more $G$'s on the bottom strand to the right of the origin again decreasing the kinetic barrier of the bonds at the origin. Thus the hydrogen bonds at the origin would have lower kinetic barriers, and hence are susceptible to break, which could underlie the origins' function. $GC$ Skew has been attributed to asymmetric mutational pressures due to the differences in leading and lagging strand replicational and trascriptional mechanisms\cite{bacterialgenomeorganization,asymmetricmutation,strandspecificmutations}, and is essentially treated as a detrimental side-effect of replicational directionality, as opposed to it being the \textit{cause} of replicational directionality as we argued above. We would like to clarify that, within our model, while the directionality of the unzipping machinery is determined by the $GC$ skew, the direction of new strand synthesis is still dictated by the $3' \hyp 5'$ orientation of the template strand, with the skew acting merely as a perturbation.

We do not preclude the possibility a self-reinforcing loop of asymmetric mutational pressure causing $GC$ skew and thus replicational directionality, which in turn causing asymmetric mutational pressure and so on. Such a self-reinforcing loop can maintain and alter, at evolutionary timescales, the amount of skew in response to environmental pressures to replicate faster or slower. Such strand-asymmetric mutational pressure associated with replicational direction has been documented\cite{mutationneighborhood,genereplicationorientation}. It is interesting to note that transcription too has been shown to lead to asymmetric mutation, selectively enriching the strand along the direction of transcription with more $G$, even when that direction is anti-parallel to that of replication due to the location of the genes on the lagging strand\cite{strandspecificmutations,transcriptionmutationasymmetry}. Cellular transcriptional machinery can take advantage of lower kinetic barrier provided by $G$-enrichment of leading strands when the genes are placed on the leading strand. Whereas, with the genes on the lagging strand, transcriptional machinery must work against the higher kinetic barrier due to the ``wrong'' $GC$ orientation. The preferential placement of genes on the leading strand, observed in genomes of multiple species\cite{strandbiasedgene}, could thus have a kinetic basis. Also, the correlation between the magnitude of skew in a genome and its replicational speed too has been documented\cite{growthrategcskew}. The counter-pressure that limits the magnitude of $GC$ skew is the obvious reduction it causes in coding possibilities, which places restrictions on the Genetic Code. Maximal $GC$ skew will significantly reduce the number of amino acids that can be encoded, with only three nucleotides available per strand (assuming $AT$  distribution across strands is not skewed), as opposed to four used in the genetic code. If not for the utility of $GC$ skew in determining the directionality of movement of unzipping machinery, it would have been selected out due to the above-mentioned counter-pressure.

The other pair of nucleotides, $A$ and $T$, are also observed to be asymmetrically distributed across the two strands of DNA in various genomes, and its switch is correlated with replicational origins\cite{yeastreplicationorigins}. But the base-pair orientation does not consistently correlate with the direction of replication across genomes of different organisms\cite{growthrategcskew}, like that of the $GC$  base-pair. For example, $T$ is enriched on the leading strand in Human genome, whereas $A$ is enriched on the leading strand in B. Subtilis genome. It is unclear whether different environmental factors or growth rates of various organisms dictate the asymmetric cooperativity mode of  the $AT$ base-pair.


A rich source of experimental evidence for the above claims, apart from the strong evidence from incorporation kinetics\cite{incorporationkinetics} presented earlier, is the documented asymmetric (polar) and sequence-dependent rate of movement of the ``unzipping machinery'', or the replicational fork (as it is called in the Biological literature) as it traverses the genome during replication. During DNA replication, the replicational fork moves unidirectionally from the origin of replication, with direction dictated by the $GC$ skew, which we explain to be due to asymmetric kinetic influence of $5' \hyp C \hyp 3'/3' \hyp G \hyp 5'$ base-pair on its neighboring base-pairs. Thus, stretches of genome with $G$-enriched on one strand should allow the fork to proceed in one direction, while inhibiting its movement in the opposite direction. Such polar inhibition of replicational forks through $G$-enriched sections has been observed\cite{zdnaorientation,trancriptionasymmetry,flipskewstopreplication}, and are usually explained as due to triple-helix formation. This sequence and orientation-dependent movement of replicational fork can be explained using the asymmetric kinetics of (un)zipping of the asymmetrically cooperative DNA. It has to be noted that the permissive and blocking orientations set by $GC$  skew are consistent for the movement of both the DNA unzipping machinery and the replicational and transcriptional machinery through $G$-enriched sections of different genomes. Thermodynamic parameters of DNA unzipping alone cannot capture such direction-dependent rates of movement of the replicational fork. More evidence for sequence- and orientation-dependent kinetics of the replicational fork are a) the orientation-dependent slowdown of the replicational fork at transcription-start and stop elements\cite{transcriptionreplication}, b) the orientation-dependent pause or termination of replication at \textit{ter} elements of E.Coli\cite{tusterpaperkornberg}, with the choice between pause and termination determined by the speed of the replisome\cite{tusterspeed},  and c) genetically-determined replicational slow zones in budding yeast\cite{replicationslowzone} and D. Melanogaster\cite{replicationslowzones} genomes. At the single-molecule level, the orientation of the terminal base-pair of DNA hair-pin molecules has been discerned using kinetics of unzipping through a nanopore\cite{orientationdiscrimination}. More recently, the differences in lifetimes of stacking interactions between swapped-sequence pairs such as $5' \hyp AT \hyp 3'/3' \hyp GC \hyp 5'$ and $5' \hyp GC \hyp 3'/3' \hyp AT \hyp 5'$ have been shown to span sveral orders of magnitude\cite{stackingforcesScience}, further supporting our hypothesis of the connection between base-pair orientation and kinetics.

\subsubsection*{Quadruplet alphabet and triplet codons}

 Within the above hypothesis, that anti-parallel heteropolymers employ asymmetric base composition to set local unzipping directionality, we can deduce that, if composed of merely two types of monomers,  DNA would have reduced information-encoding ability. To demonstrate this, let us hypothetically restrict the DNA sequences to be made of only two monomers, $G$ and $C$. Let us also assume that a nucleotide-string of length $s$ is needed to encode information about a protein. If the nucleotide sequence encoding for the protein is $5' \hyp (G)_s \hyp 3'/3' \hyp (C)_s \hyp 5'$ and the origin of replication and transcription is located to the right of this sequence (in accordance with our assumption above), then the transcription and replication will proceed smoothly. For any other encoding, such as $5' \hyp (G)_m (C)_n (G)_l \hyp 3'/3' \hyp (C)_m (G)_n (C)_l \hyp 5'$,  with $s=m+n+l$, the replication and transcription will stall in both directions for sufficiently large $n$, as has been demonstrated experimentally\cite{flipskewstopreplication}. This is due to the change in the mode of asymmetric cooperativity, within our picture. such drastic dependence of the replicational potential of a sequence on the information contained in it is detrimental to the organism, since it severely restricts the coding space available to encode information about making proteins. This also would lead to heteropolymers with minimal information content (heteropolymer single strands dominated by a single monomer type) to be replicationally more successful, leading to reduced evolvability. Fig. \ref{explanations}(e) illustrates the destruction of cooperative behavior for an arbitrary sequence. The ability to encode information (or alternately to support neutral variations on which evolution could act upon) could have been revived through the introduction of another pair of (distinct) monomers, thus resulting in four bases that we observe today\cite{fourlettercode}, as shown in Fig. \ref{explanations}(e)-(f). The four-letter code delinks replicational potential of a sequence from its information content, thus expanding the coding space.

Also, since the thermodynamic and kinetic characteristics of any given hydrogen bond in a DNA sequence is assumed to depend sensitively on its immediate neighbors, any mechanism that translates the genetic code utilizing those characteristics must also include the neighborhood. Thus, it is straightforward to rationalize the usage of three nucleotides in a codon, as it is the minimum length required for the thermodynamic and kinetic characterization of the central hydrogen bond, assuming our asymmetric cooperativity model holds for RNA as well.  It follows that the central hydrogen bond or the second base in the tri-nucleotide codon should be more important, compared to the first or the third base position, when it comes to choosing a specific genetic code among the multitude of choices. It has been shown that this is indeed the case, and the natural genetic code is chosen to tolerate mutations at first and third positions, whereas major differences between amino acids were specified by the second base in a codon\cite{secondbaseimportant,oneinamillion}. This also obviates the need for using the number of distinct amino acids, $20 \hyp 22$, hitherto an unexplained number, to explain the emergence of the triplet code.

\subsubsection*{Palindromes and Inverted Repeats}

 Consider the sequence $5' \hyp CTAG \hyp 3'/3' \hyp GATC \hyp 5'$, which has been shown to be extremely rare in bacterial genomes\cite{dinucleotidefreqpnas}. The sequence of the bottom strand can be seen to be the reverse of the top strand sequence, exhibiting a special kind of symmetry called ``dyad symmetry''. Such symmetric sequences are called palindromes. Perfect palindromes are generally under-represented in most genomes\cite{palindromeinstability}. Inverted repeats are sequences with an intervening sequence between the two symmetric ``arms'' of a palindromic sequence. As with the larger-scale approximate dyadic symmetry of the $GC$-skew-switching locations leading to origins of replication, these smaller-scale symmetry elements too serve as origins of replication and transcription\cite{palindromereplication}, and function as targets for restriction enzymes\cite{palindromerestriction}. Within our model, these properties follow from the increased symmetry of palindromic and inverted repeat sequences. The two middle hydrogen bonds in the example sequence, $5' \hyp TA \hyp 3'/3' \hyp AT \hyp 5'$, cannot influence each other asymmetrically, that is, one bond inhibiting the formation of the other and in turn be catalyzed by it. If the $5' \hyp T \hyp 3'/3' \hyp A \hyp 5'$ hydrogen bond inhibits the bond to its right, then $5' \hyp A \hyp 3'/3' \hyp T \hyp 5'$ must inhibit the bond to its left, resulting in both bonds influencing each other symmetrically. This dyadic symmetry of the palindromic arrangement excludes the presence of asymmetric cooperativity, leading to an increase in the kinetic barrier for hydrogen-bond breaking of the both base-pairs. This argument holds for $5' \hyp CG \hyp 3'/3' \hyp GC \hyp 5'$ as well. Depending upon the local directionality, one of the neighboring base-pairs of such dinucleotide sequences will have its barrier reduced from both left and right, resulting in local unzipping. The converse is also true, that of a reduction in the kinetic barrier for hydrogen-bond breaking between $5' \hyp AT \hyp 3'/3' \hyp TA \hyp 5'$ and $5' \hyp GC \hyp 3'/3' \hyp CG \hyp 5'$, as shown in Fig. \ref{explanations}(g)-(h). This results in one or more bonds weaker than the rest in the neighborhood, leaving such sequences susceptible to adopt single-stranded configurations due to bond-breaking and bidirectional cooperative unzipping, and resulting in secondary structures such as cruciforms and hairpins. More over, due to switching of asymmetric cooperativity modes at one of the two ends and at the center of a sufficiently long palindrome, DNA replication is hindered, resulting in DNA fragility at these points\cite{irunstable}.

Support for our claim above comes from its ability to collectively explain multiple apparently unrelated observations concerning the nature of sequences at genomic replicational and transcriptional origins. Local hydrogen-bond weakening, a result of the dyadic-symmetry present in palindromes, inverted repeats , $5' \hyp CG \hyp 3'/3' \hyp GC \hyp 5'$-containing CpG islands and $5' \hyp TA \hyp 3'/3' \hyp AT \hyp 5'$-containing $AT$ -rich sequences, leads to their functioning as promoters and origins of replication\cite{promoterarchitectureNature,palindromereplication,cpgislandpromoters,replicationorginsreview,transcriptionreplicationconnected}. This weakening also promotes double-strand instability\cite{invertedrepeatsunstable}, which could lead to entropy-mediated weakening of intra-strand covalent bonds, and could also explain the functioning of such symmetric sequences as recombination hotspots\cite{invertedrepeatsunstable,invertedrepeatsunstable2,invertedrepeatsstructuralevolution}.

Furthermore, the replicational stability of the trinucleotide repeats $5' \hyp (CTG)_n \hyp 3'/3' \hyp (GAC)_n \hyp 5'$ and $5' \hyp (CGG)_n \hyp 3'/3' \hyp (GCC)_n \hyp 5'$ was observed to depend on the orientation of the repeat with respect to the replicational origin\cite{ctgdirectionality,trinucleotiderepeatsbad}, providing a strong evidence for the picture we developed above. Within our model, the lack of directionality of the dinucleotide sequence $5' \hyp GC \hyp 3'/3' \hyp CG \hyp 5'$ in the trinucleotide repeats would leave the $T/A$  and $G/C$ base-pair to dictate the directionality of replicational machinery movement. Thus, $5' \hyp CTG \hyp 3'/3' \hyp GAC \hyp 5'$ would allow the replicational machinery to pass through only when approached from the permissive direction due to asymmetric cooperativity, whereas $5' \hyp CAG \hyp 3'/3' \hyp GTC \hyp 5'$ would block the machinery, inducing errors in replication and resulting in instability. The dependence of instability or fragility on the repeat length of various types of repeats suggests the presence of cooperativity in such sequences. The cooperativity of such sequences would manifest in extension or deletion of the repeat units in multiples than as single units, as has been experimentally observed\cite{ctgdirectionality}. Also, as we have mentioned above, the mutual strengthening of kinetic barrier in $5' \hyp CG \hyp 3'/3' \hyp GC \hyp 5'$ and $5' \hyp TA \hyp 3'/3' \hyp AT \hyp 5'$ and the mutual weakening of the barrier in $5' \hyp GC \hyp 3'/3' \hyp CG \hyp 5'$ and $5' \hyp AT \hyp 3'/3' \hyp TA \hyp 5'$ would lead to significant difference in the mobility of replicational machinery through these two types of dinucleotide sequences, with the sequences with higher barrier hindering the mobility. This would result in the selective advantage and relative enrichment of one over another over the entire genome. This genome-wide relative enrichment of $GC$ and $AT$ over $CG$ and $TA$ dinucleotide sequences has been documented across the genomes of multiple species\cite{comparativegenomics,dinucleotidesignature,dinucleotideasymmetry,signalprocessingpaper}.

\subsubsection*{Helicity}

 Thus far, we have elaborated on the replicational benefits of asymmetric cooperativity and its instantiation in DNA through heteromolecular base-pairs. The orientation-dependent kinetics of replication and transcription depended on the orientation of the base-pairs, with the latter determining the local asymmetric cooperativity mode. Alterations of these orientations (and hence of the sequence) through mutations, such as translocations, on evolutionary timescales, provided the variations in the kinetic behavior of the genome on which evolution could act upon, thereby increasing the fitness of the species. Similarly, some amount of control on the kinetics of replication and transcription, \textit{on the timescale of the lifetime of an individual organism}, would have proved very useful for the organism in order to respond to the changes in its environment, and would have helped its acclimatization. Control of asymmetric cooperativity, and through it the kinetics of replication and transcription, can be effected at shorter timescales if asymmetric cooperativity is made to depend on the more malleable local \textit{structure} of the DNA, in addition to it being dependent on the \textit{sequence}. We argue that helicity of DNA provides exactly such a capability.

The opposite angular displacements of neighboring hydrogen bonds to the left and right of a given hydrogen bond, due to helicity, could also impose opposing kinetic influences on the bonding and dissociation of the central hydrogen bond, with the direction of inhibition and catalysis dependent on the orientation of the central hydrogen bond. A local reduction in helicity, and hence a reduction in the opposite angular displacements of neighboring hydrogen bonds, would symmetrize the structure and thus would reduce asymmetric cooperativity. Such locally decreased asymmetry in kinetic influence over a hydrogen bond will weaken it and/or its neighbor, and would result in local weakening and unwinding of the double-strand, just as the dyadic symmetry of $GC$ skew-switching and palindromic sequences weaken the double-strand. This helicity-dependence of asymmetric cooperativity could offer an ``epigenetic'' route to regulate origins of replication and transcription. The following lines of evidence support this picture.

It is an experimentally well-established fact\cite{topologygenecontrol,topologyreview1,torquesdnaopening,negativesuperhelicityOric2} that DNA negative superhelicity, which can be defined as the reduction in the angle between two successive base-pairs in the double helical structure, weakens the inter-strand hydrogen bonds and locally unwinds DNA at specific sequences. In one of the first demonstrations, a promoter of transcription in E. Coli was observed to locally unwind, specifically at $3' \hyp TA \hyp 5'/5' \hyp AT \hyp 3'$, under negative superhelical stress\cite{negsupercoilpromoter}. In another experiment, superhelically constrained circular DNA strands were shown to adopt single-stranded configuration at inverted repeat locations, whereas the same strands, when linearized, did not show such instability\cite{circlinearnegsupercoil}. In one of the more recent experiments\cite{negsupercoilsinglemolecule}, local unwinding at the origins of replication of single molecule double-strand DNA under negative superhelical stress were visually observed. Negative superhelical stress has been shown to destabilize DNA double strands near termination and promoter regions of plasmids and viral genomes\cite{negsupercoilreplicationorigin}. It has also been theoretically noted that a collective twisting of the DNA strand is required to nucleate local denaturation and bubble formation of DNA\cite{localtwisthelicity}. Explanation for such behavior is straight-forward within our model, where, the already weakened inter-strand bond(s) at these dyadic-symmetric sequences is further destabilized due to the reduction in asymmetric cooperativity caused by negative superhelicity, leading to their local unwinding.

The normal form of DNA is a right-handed helix, denoted as the B-form, whereas, the Z-form, which has specific sequence requirements and predominantly occurs in negatively superhelical genomes, is left-handed\cite{energeticsBZtransition}. The interface between the B-DNA and Z-DNA exhibits significant unwinding\cite{energeticsBZtransition}. This local loss of helicity, and thus the decreased directional asymmetry, at the B-Z interface results in weakening of the hydrogen bonds and the interface has been shown to adopt single-stranded configuration\cite{BZsinglestrand}. This interface also has been shown to stop transcription only when the DNA is negatively supercoiled\cite{zdnastopsreplication}, understandable from the asymmetric cooperativity model as being due to change in the mode of asymmetric cooperativity, due to the change in the sign of helicity at the B-Z interface. If the (un)zipping kinetics is dependent upon helicity of DNA through asymmetric cooperativity, as claimed above, then global DNA supercoiling structure, by influencing local helicity, can govern transcription and replication, as has been argued here\cite{supercoilinghomeostasis,topologygenecontrol,negativesuperhelicityOric2}. Transcription and replication can thus respond to environmental changes, since the latter is known to influence DNA supercoiling\cite{superhelicityenvironment,topologyreview2}. As an interesting example, the global topological state of cyanobacterial genome has been shown to be strongly correlated with the circadian gene expression state\cite{oscillationsgeneexpression}.

Helicity, which structurally instantiates asymmetric cooperativity in DNA and thus enables topological control of replication and transcription as argued above, also imposes structural constraints on the monomers constituting the DNA double helix. The ground-state double helical structure of DNA is patently a three-dimensional one, which implies that the covalent and hydrogen bonds that attach a nucleotide monomer to its neighbors cannot all be in the same plane, i.e., the bonds cannot be coplanar. If the bonds were all in the same plane, then the ground state structure of DNA will be that of a flat untwisted ribbon, not a double helix. We will need this constraint below, to explain the chirality of nucleotides.

\clearpage
\setcounter{figure}{5} 
\begin{figure}
{\raisebox{-0.8\height}{\includegraphics[ scale = 0.053, keepaspectratio]{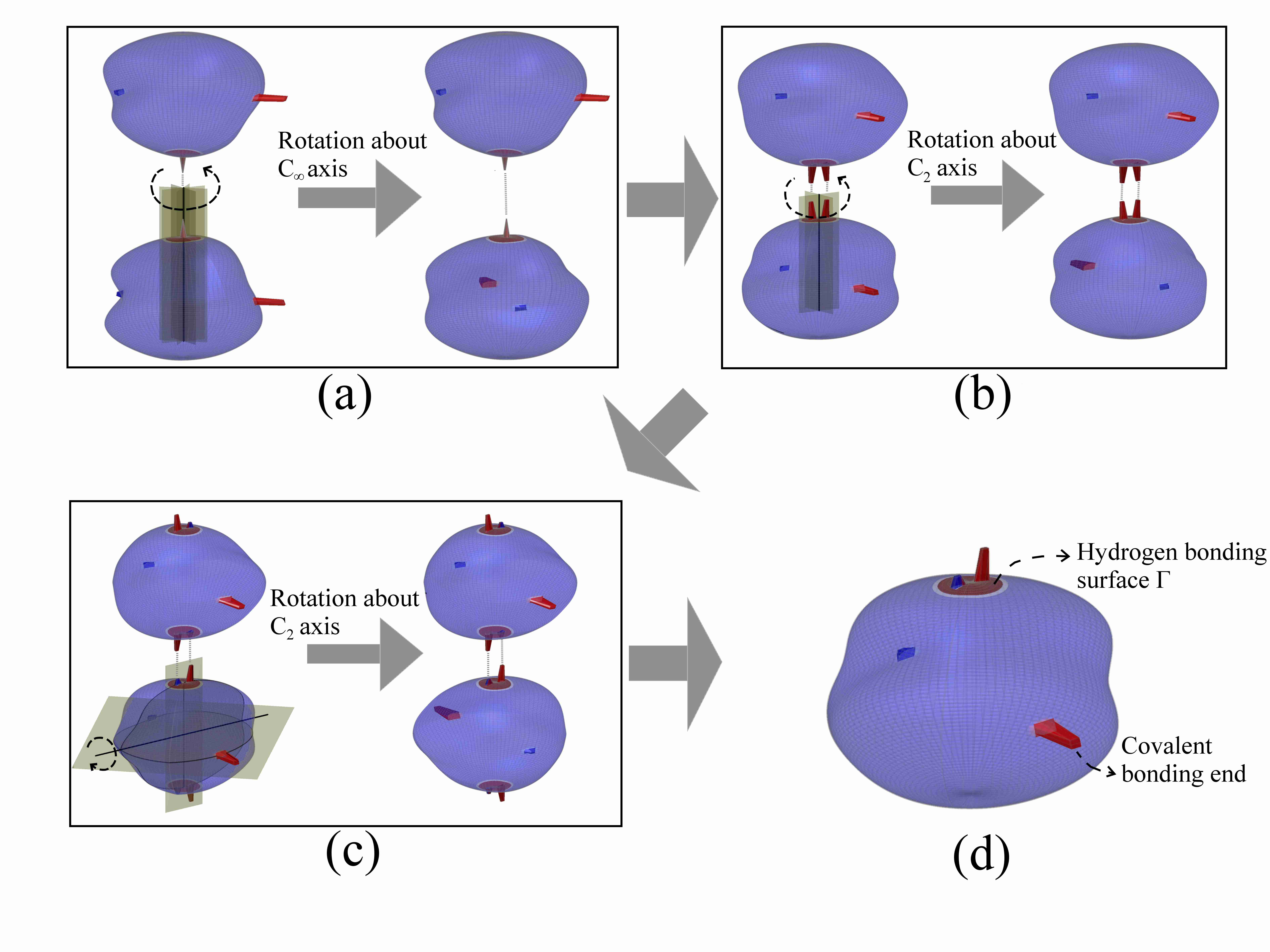}}}
\caption{An illustration of the relationsip between reflection and rotational symmetry operations, and its relevance to the relationship between chirality and orientational specificity. The latter is a requirement that monomers must satisfy during hydrogen bonding, for efficient heteropolymer self-replication. A monomer is represented here as an approximate spheroid, with protrusions near equator denoting covalent bonding ends and the ones in poles denoting hydrogen bonding ends. Local and global symmetry planes are represented by greenish gray planes, and rotational symmetry axes, by thick black lines. Non-coplanarity of all the bonds assumed. (a) Illustration of the lack of orientational specificity during hydrogen bonding between two monomers with just one hydrogen bonding end each within the surface $\Gamma$. An infinite number of local non-parallel reflection symmetry ($\sigma$) planes exist within $\Gamma$, with the line of intersection between the planes being the rotational symmetry axis, $C_{\infty}$. (b) Two similar hydrogen bonding ends support two local reflection symmetry planes within $\Gamma$, still providing a $C_2$ rotational symmetry axis. (c) Two dissimilar hydrogen bonding ends within $\Gamma$ provides one reflection symmetry plane. The presence of another global or local non-parallel reflection symmetry plane, such as the one shown across the equator, would introduce a rotational symmetry axis, destroying orientational specificity. (d) For monomers with not more than two hydrogen bonds within a single $\Gamma$ surface, the imposition of orientational specificity leads to chiral structure, as long as non-coplanarity of all the bonds are assumed. }
\label{explainchirality}
\end{figure}

\subsubsection*{Chirality}

One of the longstanding puzzles in Biology is the same-handedness or homochirality of nucleotides and amino acids. Nucleotides, the monomers that constitute DNA, are right-handed, whereas amino acids, monomers of proteins, are left-handed. What evolutionary advantage accrues to living systems from homochirality is unknown. But a demonstrated disadvantage of homochirality is the cross-inhibition of polymerization when both left- and right-handed enantiomers are present in the primordial growth medium\cite{PolymerizationOrientationDependence,chiralpurity,enantiomericcrossinhibition}. Then, the question of homochirality can be resolved into two related but distinct questions: a) Why did molecular evolution not choose achiral monomers to construct genetic polymers? b) In case there is an advantage in choosing chiral monomers, how did one enantiomer come to dominate in living systems? The tentative answers that have been presented so far in the literature address only the second question, where, analyses usually start with a racemic mixture of chiral molecules and proceed to identify a symmetry-breaking mechanism that chooses one enantiomer over another (reviewed in \cite{chiralityreview}). This includes the interesting explanation that the presence of significant (combinatorial) entropic barrier in forming specific structures in a racemic medium\cite{entropychirality} results in living systems prefering homochiral media. It is entirely logical to ask the first question, that \textit{why achiral molecules are disfavored}, particularly within our premise that an infinite variety of precursor molecules existed for prebiotic evolution to select an appropriate monomer. Below, we answer both the questions, by showing that, within the constraints of directionality and helicity, it is evolutionarily advantageous for monomers to be homochiral. 

We would like to recall our evolution-based explanation for directionality, above, where we argued that the instantiation of asymmetric cooperativity requires monomers that are asymmetric along the covalent-bonding direction. Thus, if asymmetric cooperativity mode has to be preserved, the \textit{5' \hyp 3' orientation} of the nucleotide monomers during polymerization must be maintained. The biochemical evolution that preceded and resulted in DNA/RNA would have had to choose a structure that imposed such an \textit{orientational restriction}, by disallowing hydrogen bonding in all but one orientation\cite{PolymerizationOrientationDependence}. Symmetry elements in the monomer structure, by providing more orientational possibilities during hydrogen bonding, would increase the chances of replicational failures by allowing hydrogen bonding with wrong covalent bond orientation, and hence would have been selected out during prebiotic evolution (see Fig. $6$). We argue below that it is best to use chiral monomers to impose orientational specificity during hydrogen bonding.  

The general idea we use below to demonstrate the need for chirality is quite simple. If we aspire to attach two objects together in just one way, and in only one orientation, using only the objects' shape complementarity, much like Lego blocks, we need to \textit{reduce the structural symmetries of the objects}. Thus, the requirement of orientational specificity imposes constraints on the presence of rotational and reflectional symmetries in the object. 

Let us represent the monomer as a general three-dimensional object, with an arrow embedded in it denoting the direction of the covalent-bonding axis $\vec{V}$ (see Fig. $6$). Let us assume that the hydrogen bonding ends lie within a small area $\Gamma$, on the surface of the object, somewhere between the two ends of the arrow. The presence of a directional covalent bonding axis results in a drastic reduction of possible symmetry elements in the monomer. Adopting the definition of chirality as the absence of improper rotation axes $S_n$ of any order $n$ (which includes reflection symmetry planes and inversion centers as special cases), it is obvious that $S_n$ axes of all orders that have a component along the covalent bonding axis $\vec{V}$ must be absent, for, they invariably switch the direction of $\vec{V}$ during reflection. Also, $S_n$ axes that are perpendicular to $\vec{V}$ and are of order greater than one ($n>1$) must be absent, since only a full $2\pi$ rotation will bring the $\vec{V}$ back to coincide with itself. Thus, we are left only with the possibility of existence of reflection planes (henceforth denoted by $\sigma$) parallel to $\vec{V}$.

In order to eliminate the possibility of existence of these reflection planes, we use a known relationship between reflection and rotational symmetry elements. The existence of two non-parallel reflection symmetry planes in the monomer, $\sigma_{\vec{m}}$ and $\sigma_{\vec{n}}$, intersecting at an angle $\phi$, implies the existence of rotational symmetry element $C_n$ of order $n = 2\pi/2\phi$, about an axis $I(\vec{m}, \vec{n})$ defined by the line of intersection of the two planes\cite{chemicalapplications}. Thus, if the monomer has at least two such $\sigma$ planes with an intersection axis $I$, we are assured of the existence of a $C_n(I)$ axis that would enable the monomer to hydrogen bond with more than one covalent bonding axis orientation, as long as $I$ and $\vec{V}$ are not coincidental. Since such multiple orientational possibilities are detrimental to successful self-replication, monomers with multiple $\sigma$ planes would have been selected out. Since the orientational possibilities during hydrogen bonding are dictated more by \textit{local} symmetry elements within $\Gamma$, with global symmetry elements or lack thereof merely perturbing the bonding energies (and are usually not strong enough to prevent bonding outright \cite{hbondorientation}), we need to include the approximate local symmetry elements within $\Gamma$ as well in our evaluation of orientational possibilities during monomer bonding. 

The case of a single hydrogen bonding end within $\Gamma$ can thus be excluded, because of the existence of a local $C_{\infty}$ axis, which allows for hydrogen bonding with more than one possible $\vec{V}$ orientation (see Fig. $6$(a)). The case of two similar hydrogen bonding ends within $\Gamma$ can also be excluded, since the arrangement has two local $\sigma$ planes, making both parallel and anti-parallel covalent bonding possible (Fig. $6$(b)). Two dissimilar hydrogen bonding ends within $\Gamma$ still has a local $\sigma$ plane, as the plane that contains both the bonding ends (Fig. $6$(c)). Thus, there cannot be any more $\sigma$ planes, local or global, as their existence would imply the existence of a $C_n$ axis and hence orientational degeneracy, as long as the local $\sigma$ plane does not contain $\vec{V}$.  Thus, the monomer is either chiral, or, the two hydrogen bonds within $\Gamma$ and the two covalent bonding ends are coplanar. The latter case, where the local $\sigma$ plane constituted by the two hydrogen bonds also contains $\vec{V}$, is expressly prohibited, because such a coplanar configuration, with all the bonds (covalent and hydrogen bonds) lying in the same plane, cannot lead to a three-dimensional helical structure, as argued in the ``Helicity'' section.

These arguments provide the rationale for the evolutionary choice of chiral biological macromolecules, with the choice of a specific enantiomer determined by small initial random fluctuations\cite{SoaiReaction}. One way of verifying our arguments above for the need for chirality is to introduce reflection-symmetry planes in the nucleotide structure, without destroying its $3' \hyp 5'$  directionality. It becomes obvious that the resultant structures would possess enough symmetry to base-pair with another nucleotide in opposite orientation, as illustrated in figure $6$. We are aware of at least one other proposal connecting directionality and chirality, but on the basis of a completely different line of reasoning\cite{directionalitychirality}.

As for the second question, the presence of monomers of opposite chirality in a racemic primordial environment would have severely inhibited self-replication of chiral heteropolymers, since the former would hydrogen bond with wrong $\vec{V}$ orientation, thereby altering the mode of asymmetric cooperativity, and thus strand elongation. Evidence for the above argument comes from this experiment\cite{PolymerizationOrientationDependence}, where monomers of opposite chirality were shown to be incorporated as chain terminators during double strand formation. Thus asymmetric cooperativity requires homochirality for unhindered strand growth. This enantiomeric cross-inhibition necessitates the presence of a cross-chiral catalyst that would help both in the supply of enantiopure monomers and in orienting the monomers for hydrogen bonding, thus reducing the significant orientational entropy contribution to the kinetic barrier. We could envisage, as an instantiation of Eigen's hypercycle\cite{eigenhypercycle}, a protolife that coupled two self-replicating heteropolymer systems made of two different kinds of molecules with opposite chiralities (viz. nucleotides and amino acids), each catalyzing the polymerization of the other, apart from autocatalyzing their own synthesis\cite{rnaselfreplication,peptideselfreplication,peptidehypercycle,peptideselfreplication2}. Such a hypercyclic arrangement, by resolving the problem of enantiomeric cross-inhibition, would have provided further evolutionary advantage to the coupled system.

\section*{Conclusion}

We have found, in our model of self-replication of hypothetical autocatalytic heteropolymers, that unequal kinetic influence of inter-strand hydrogen bonds on their left and right neighbors improves the replicational potential substantially. This improvement is due to the simultaneous satisfaction of two competing requirements of both long lifetime of inter-strand hydrogen bonds to assist in covalent bonding, and low kinetic barrier for easy formation and dissociation of hydrogen bonds to speed up replication. This broken-symmetry mechanism is shown to lead to strand directionality, and through the consequent requirement of orientational specificity of a three-dimensional structure, to monomer homochirality. The derivation of latter can be motivated by a simple analogy: to restrict two objects to attach to each other only in one specific orientation requires reduction in the structural symmetry of the objects. Surprisingly, this symmetry argument even provides an explanation for the non-existence of single hydrogen bonds for base-pairing interaction in DNA.

The presence of asymmetric cooperativity in DNA, strongly suggested by multiple lines of experimental evidence listed above, provides an unifying explanation for a number of structural and functional elements of DNA. Sequence-dependence of asymmetric cooperativity, and hence unzipping directionality, in anti-parallel strands makes the latter evolutionarily superior over parallel strands with frozen-in directionality, by enabling acquisition of multiple origins of replication. Within anti-parallel strand configuration, incorporation of asymmetric cooperativity requires breaking of compositional symmetry between the two strands. This leads to heteromolecular base-pairing, and asymmetric base composition determines local asymmetric cooperativity modes. The reduction in coding possibilities in a compositionally asymmetric double-strand with only two types of monomers is solved by using four nucleotide alphabets. The need to include immediate neighborhood in the kinetic and thermodynamic characterization of a hydrogen bond explains triplet genetic code.

Any local symmetrization, resulting from $GC$ skew switching across strands, $AT$ -rich sequences, CpG islands, palindromes, inverted repeats, interface between right- and left-handed helical DNA, or negative supercoiling, leads to weakening of inter-strand hydrogen bonds and double-strand instability, and thus can serve as origins or termination points of replication and transcription. These multiple types of local symmetry elements possibly function at distinct spatio-temporal and energetic scales. Helical structure of DNA, by structurally instantiating asymmetric cooperativity, could connect the DNA's global positive or negative supercoiling stress, a function of the physico-chemical environment in the cell, to modulation of local structural asymmetry, which governs transcription and replication. We also speculate that the kinetics of unzipping underlie information-encoding mechanisms in genomes, with thermodynamics playing a more subdued role.

In aiming to provide simple, broad-brush explanations of the structure and function of DNA in the light of asymmetric cooperativity, we ignored the contributions of proteins in regulating replicational and transcriptional mechanisms. When enzymes and proteins perform functions on the DNA by spending energy through hydrolysis of ATP, they can transcend the kinetic limitations imposed by asymmetric cooperativity of DNA. This limits our ability to predict the direction of DNA-based reactions with certainty, and the above picture only helps identify ``propensities'', the direction reactions would take in the absence of energetic input.

In conclusion, we have built a very simple model of primordial heteropolymers to show that asymmetric cooperativity improves replicational potential, and utilized it to tentatively explain multiple fundamental properties of DNA. We have made an experimentally verifiable prediction to test our hypothesis of asymmetric cooperativity in DNA. We have identified a single property, dyadic symmetry, that underlies the diverse sequences that function as origins of replication and transcription, and connected them with other fundamental properties of DNA, such as heteromolecular base-pairing and asymmetric nucleotide composition. We have identified an evolutionary advantage for strand directionality and anti-parallel orientation of DNA double-strand, thereby providing a rationale for the existence of a complicated lagging strand replicational machinery. We have argued that asymmetric nucleotide composition or $GC$ skew sets the replication orientation, which also leads to quadruplet alphabet of genome. Finally, we have shown that chirality of nucleotides ensures orientational specificity during polymerization. The model's explanatory range, unifying hitherto unconnected properties of DNA, with minimal assumptions, provides us confidence in its correctness. 


\subsection*{Acknowledgements:} 
Support for this work was provided by the Moffitt Physical Science and Oncology Network (PS-ON) NIH grant, $U54CA193489$.
 
We thank John Cleveland, Joel Brown and Robert Gillies for useful comments. HS thanks IMO faculty, Artem Kaznatcheev and other post-doctoral associates for helpful discussions.



\clearpage

\printbibliography

\section*{Statement of Author Contribution}
RG and HS conceived the topic of research. HS modeled, analyzed the results and arrived at the explanations. HS and RG co-wrote the paper. 

\section*{Statement of Competing Financial Interests}
The authors declare no competing financial interests.

\end{document}